\documentclass[
superscriptaddress,
amsmath,amssymb,twocolumn,
aps,pra,
]{revtex4-2}

\usepackage{graphicx}
\usepackage{dcolumn}
\usepackage{bm}
\usepackage{physics}
\usepackage{amsmath}
\usepackage{hyperref}
\graphicspath{{Images/}}

\newcommand{\tucadd}{School of Electrical and Computer Engineering, Technical University of Crete, Chania, Greece 73100}
\newcommand{\angelqadd}{AngelQ Quantum Computing, 531A Upper Cross Street, $\#$04-95 Hong Lim Complex, Singapore 051531}

\begin{document}

\preprint{APS/123-QED}

\title{Fock State-enhanced Expressivity of Quantum Machine Learning Models}

\author{Beng Yee Gan}
  \email{gan.bengyee@u.nus.edu}
  \affiliation{Centre for Quantum Technologies, National University of Singapore, 3 Science Drive 2, Singapore 117543}

\author{Daniel Leykam}
  \affiliation{Centre for Quantum Technologies, National University of Singapore, 3 Science Drive 2, Singapore 117543}
\author{Dimitris G. Angelakis}
 \email{dimitris.angelakis@gmail.com}
 \affiliation{Centre for Quantum Technologies, National University of Singapore, 3 Science Drive 2, Singapore 117543}%
 \affiliation{\tucadd}%
 \affiliation{\angelqadd}

\date{\today}

\begin{abstract}
The data-embedding process is one of the bottlenecks of quantum machine learning, potentially negating any quantum speedups. In light of this, more effective data-encoding strategies are necessary. We propose a photonic-based bosonic data-encoding scheme that embeds classical data points using fewer encoding layers and circumventing the need for nonlinear optical components by mapping the data points into the high-dimensional Fock space. The expressive power of the circuit can be controlled via the number of input photons.  Our work shed some light on the unique advantages offered by quantum photonics on the expressive power of quantum machine learning models. By leveraging the photon-number dependent expressive power, we propose three different noisy intermediate-scale quantum-compatible binary classification methods with different scaling of required resources suitable for different supervised classification tasks. 
\end{abstract}

\maketitle

\section{Introduction}

Machine learning approaches such as artificial neural networks are  powerful tools for solving a wide range of problems including image classification and regression. However, the scalability of machine learning implemented using general-purpose electronic circuits is limited by their high power consumption and the end of Moore's law. These issues motivate the pursuit of dedicated hardware for machine learning including photonic neural networks~\cite{de2019photonic,hamerly2019large,roques2020heuristic,shastri2021photonics} and quantum circuits~\cite{peruzzo2014variational,mitarai2018quantum,benedetti2019parameterized,schuld2020circuit, fujii2020quantum, goto2021universal,lloyd2020quantum}. 

The combination of ideas from the photonic and quantum machine learning communities may enable further speed-ups and novel functionalities \cite{chatterjee2016generalized,schuld2019quantum,steinbrecher2019quantum,killoran2019continuous,bartkiewicz2020experimental,taballione202012,chabaud2021quantum,ghobadi2021nonclassical}. For example, both classical and quantum photonic neural networks are presently limited by the difficulty of incorporating nonlinear activation functions. This challenge can be circumvented using the kernel trick, in which the input data is mapped into a high-dimensional feature space where simple linear models become effective \cite{schuld2019quantum,havlivcek2019supervised,schuld2021machine,schuld2021supervised}. The simplest quantum feature map based on repeated application of data-dependent single qubit rotations is already sufficient to serve as a universal function approximator~\cite{perez2020data,schuld2021effect,perez2021one}.

Despite progress in various aspects of near-term quantum machine learning algorithms \cite{li2022recent} including experimental realizations \cite{bartkiewicz2020experimental,dutta2021realization,kusumoto2021experimental,peters2021machine,ren2022experimental}, proposals for various platforms \cite{kusumoto2021experimental,tangpanitanon2020expressibility} and studies of statistical properties of quantum machine learning models \cite{abbas2021power,holmes2022connecting,caro2021generalization}, the encoding of input data is still a significant bottleneck for (quantum) photonic machine learning hardware. For example, the expressive power of quantum circuits based on parameterized single qubit rotations is limited by the number of encoding gates used~\cite{perez2020data,schuld2021effect}. Similarly, some existing quantum machine learning algorithms with proven speedups for future fault-tolerant quantum computers assume the existence of quantum-random access memory (RAM) \cite{Giovannetti_2008} that can provide the input data in a quantum superposition with no overhead \cite{harrow2009quantum,wiebe2012quantum,lloyd2013quantum,lloyd2014quantum,rebentrost2014quantum,biamonte2017quantum,dunjko2018machine}. Yet, the sources of the speedup of these algorithms are still under active debate \cite{tang2021quantum,cotler2021revisiting}. Thus, a pressing goal is to develop machine learning algorithms that avoid encoding large input datasets \cite{lloyd2016quantum,harrow2020small,liu2021rigorous} or more efficient data-encoding methods. This article addresses the latter problem.

\begin{figure}[tb]
\includegraphics[width=80mm,scale=0.5]{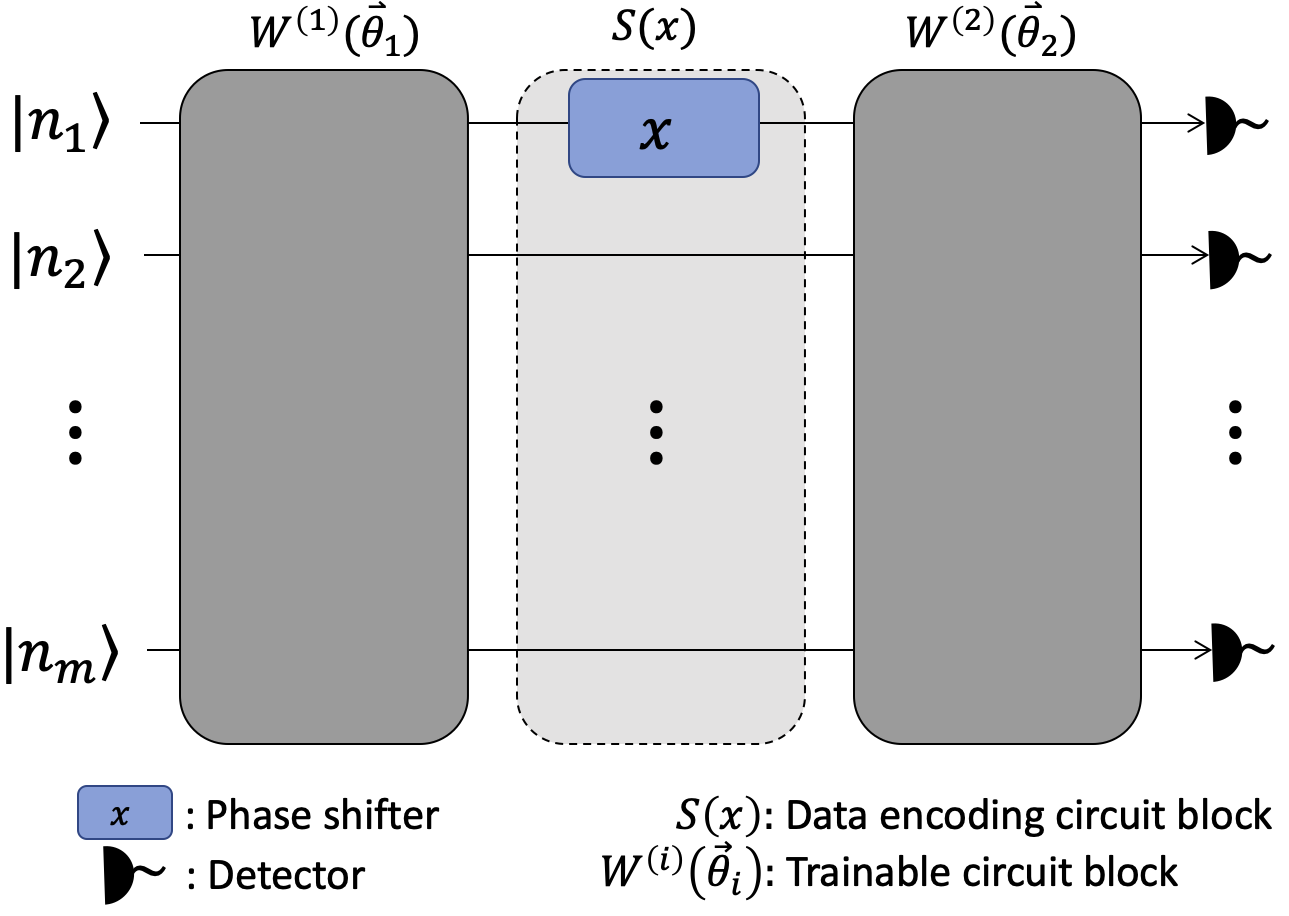}
\caption{\label{CircuitDiagram}Circuit diagram of parameterized linear quantum photonic circuit with $m$-spatial modes and encoding data $x$ using a single phase shifter. The expectation value with respect to observables of photon-number resolving (PNR) or threshold detectors can be written as a Fourier series $\sum_{\omega} c_\omega e^{i\omega x}$, with frequencies $\omega$ determined by the number of photons fed into the circuit, while the coefficients $c_\omega$ are determined by the trainable circuit blocks and the observable.}
\end{figure}

Specifically, we generalize the qubit-based circuit architecture analyzed in Refs.~\cite{perez2020data,schuld2021effect} to quantum photonic circuits (QPCs) constructed using linear optical components such as beam splitters and phase shifters, photon detectors, and Fock state inputs. We consider parameterized linear QPCs [Fig.\ref{CircuitDiagram}(a)] consisting of two trainable circuit blocks with one data encoding block sandwiched between them. We show that for a fixed number of encoding phase shifters, the expressive power of the parameterized quantum circuit is improved by embedding the classical data into the higher-dimensional Fock space. This enables the approximation of classical functions using fewer encoding layers while circumventing the need for nonlinear components. The origin of this improved encoding efficiency is that each phase shifter simultaneously uploads the input data onto multiple Fock basis states simultaneously. 

Similar to Ref.~\cite{schuld2021effect}, $n$-photon quantum machine learning models can be expressed as a Fourier series
\begin{equation}
    \begin{aligned}[b]
    f^{(n)}(x,\bm{\Theta},\bm{\lambda})= \sum_{\omega \in \Omega_n} c_{\omega}(\bm{\Theta},\bm{\lambda}) e^{i\omega x},
    \end{aligned}
\end{equation}
where $\Omega_n \in \mathbb{N}$ is the frequency spectrum and $\{c_{\omega}\}$ are the Fourier coefficients that depend on trainable circuit block's parameters $\bm{\Theta}  = (\bm{\theta}_1,\bm{\theta}_2)$  and observable's parameters $\bm{\lambda}$. The expressive power of the Fourier series is determined by two components: the spectrum of frequencies $\omega$, and the Fourier coefficients $c_{\omega}$. We show that the frequency spectrum of the circuit can be controlled by the number of input photons. Thus, a rich frequency spectrum can be generated by providing sufficient number of input photons to linear QPCs with a constant number of spatial modes. In contrast, qubit-based circuits require deeper or wider circuits to increase the size of their frequency spectrum. When generalized to arbitrary input states and observables the QPCs can also generate arbitrary set of Fourier coefficients that combine the frequency dependent basis functions $e^{iwx}$, allowing them to approximate any square-integrable function on a finite interval to arbitrary precision~\cite{carleson1966convergence,weisz2012summability,schuld2021effect}.

As an application of the parameterized linear quantum photonic circuits, we consider three different machine learning approaches for supervised data classification: (1) A variational classifier based on minimizing a cost function by training the circuit parameters. (2) Kernel methods, which employ fixed circuits, with training carried out on observables only. (3) Random kitchen sinks, which use a set of random circuits to approximate a desired kernel function. Each of these methods has different scaling with the dimension of the data and number of training points used, and so each is better-suited to different types of supervised learning problems.

The outline of this paper is as follows. Sec.~\ref{sec:model} introduces our proposed linear quantum photonic circuit architecture and analyzes how its expressive power depends on the number of spatial modes and input photons. Next, Sec.~\ref{sec:classification} illustrates the photon number-dependent performance of the circuit for supervised classification problems. Sec.~\ref{sec:conclusion} concludes the paper.

\section{\label{sec:model} Parametrized linear quantum photonic circuit model}
To demonstrate the Fock state-enhanced expressive power of linear quantum photonic circuits, in this Section we consider the encoding of univariable functions onto circuit's output. For simplicity we consider the circuit architecture illustrated schematically in Fig.~\ref{CircuitDiagram}, consisting of a single data-dependent encoding layer $\mathcal{S}$ sandwiched between two trainable beam splitter meshes $\mathcal{W}^{(1,2)}$, described by the unitary transformation
\begin{equation}
    \mathcal{U} (x,\bm{\Theta}) = \mathcal{W}^{(2)}(\bm{\theta}_2)\mathcal{S}(x)\mathcal{W}^{(1)}(\bm{\theta}_1),
\end{equation}
where $\bm{\Theta} = (\bm{\theta}_1,\bm{\theta}_2)$ parameterizes transformations applied by trainable beam splitter meshes and $x$ is the input data.  The $n$-photon quantum model (circuit's output) is defined as the expectation value of some observable $\mathcal{M}(\bm{\lambda})$ with respect to a state prepared via the parameterised linear QPC, 

\begin{equation} \label{Eqn-Qmodel}
    \begin{aligned}[b]
    f^{(n)}(x,\bm{\Theta},\bm{\lambda}) = \bra{\textbf{n}^{(i)}}\mathcal{U}^\dagger (x,\bm{\Theta})\mathcal{M}(\bm{\lambda}) \mathcal{U}(x,\bm{\Theta}) \ket{\textbf{n}^{(i)}},
\end{aligned}
\end{equation}
where $\ket{\textbf{n}^{(i)}} = \ket{n^{(i)}_1,n^{(i)}_2,\dots,n^{(i)}_m}$ is the input $n$-photon Fock state with $n = \sum_i^m n^{(i)}_i$ and $\bm{\lambda}$ parameterizes the observable. We consider measurements made using either photon number-resolving (PNR) detectors or single photon (threshold) detectors, corresponding to $\mathcal{M}$ being diagonal in the Fock state basis with $d$ or $d^{\prime}$ distinct parameterized eigenvalues $\{\lambda_j \in \bm{\lambda}\}^{d^{(\prime)}}_{j = 1}$, respectively.

The multi-mode Fock state unitary transformation $\mathcal{U}(x,\bm{\Theta})$ is constructed from permanents of submatrices of the $m$-mode linear transformation matrix $U(x,\bm{\Theta}) = W^{(2)}(\bm{\theta}_2) S(x) W^{(1)}(\bm{\theta}_1)$ using the scheme of Ref.~\cite{scheel2004permanents} with $W^{(i)}$ as the programmable transfer matrix, describing the universal multiport interferometer that realizes arbitrary linear optical input-output transformations~\cite{reck1994experimental,clements2016optimal,bell2021further}. Each trainable unitary $W^{(i)}(\bm{\theta}_i)$ is parameterized by a vector $\bm{\theta}_i$ of $m(m-1)$ phase shifter and beam splitter angles constructed using the encoding of Reck et al.~\cite{reck1994experimental}. The data encoding block $S(x)$ employs a single tunable phase shifter placed at the first spatial mode. 

\subsection{$n$-photon quantum models as Fourier series}
In this Section we will show how to express the $n$-photon quantum models as a Fourier series. For simplicity, we consider arbitrary unitary operations $\mathcal{W}(\bm{\theta}) = \mathcal{W}$, an arbitrary parameterized observable obtained using PNR detectors $\mathcal{M}(\bm{\lambda}) = \mathcal{M}$, and $\ket{\bm{n}^{(i)}} = \ket{n,0,\dots,0}$ as the initial Fock state. The component of the output quantum state with photon numbers $\bm{n}^{(f)}$ can be written as \cite{schuld2021effect}
\begin{equation}
    \bra{\bm{n}^{(f)}} \mathcal{U}(x) \ket{ \bm{n}^{(\mathrm{i})}} = \sum_{\bm{n}^{\prime}} \mathcal{W}^{(2)}_{\bm{n},\bm{n}^{\prime}} e^{i n_1^{\prime} x} \mathcal{W}^{(1)}_{\bm{n}^{\prime}, \bm{n}^{(i)}}, \label{eq4}
\end{equation}
where the summation runs over the basis of $d = \begin{pmatrix} n+m-1 \\ n \end{pmatrix}$ Fock states corresponding to different combinations of $n$ photons in the $m$ spatial modes. The data encoding block imposes a phase shift proportional to the number of photons in the first mode following the first beam splitter mesh.

The output of full model Eq.~\eqref{Eqn-Qmodel} is obtained by taking the modulus square of Eq.~\eqref{eq4}, multiplying by the corresponding observable weight, and then summing over all output Fock basis states. This yields an expression of the form
\begin{equation}
    f^{(n)}(x) = \sum_{\bm{n}^{\prime \prime},\bm{n}^{\prime}} a_{\bm{n}^{\prime \prime},\bm{n}^{\prime}}e^{i(n^\prime_1-n^{\prime \prime}_1) x},
\end{equation}
where $a_{\bm{n}^{\prime \prime},\bm{n}^{\prime}}$ contain the matrix elements from the unitaries $\mathcal{W}^{(i)}$ and measurement's observable $\mathcal{M}$,
\begin{equation}
    a_{\bm{n}^{\prime \prime},\bm{n}^{\prime}} = \sum_{\bm{n}} \mathcal{W}^{*(1)}_{\bm{n}^{(i)},\bm{n}^{\prime \prime}}\mathcal{W}^{*(2)}_{\bm{n}^{\prime \prime},\bm{n}} \mathcal{M}_{\bm{n},\bm{n}} \mathcal{W}^{(2)}_{\bm{n},\bm{n}^{\prime}} \mathcal{W}^{(1)}_{\bm{n}^{\prime},\bm{n}^{(i)}}.
\end{equation}
This expression can be simplified by grouping the basis function with the same frequency $\omega = n^\prime_1 -n^{\prime \prime}_1$. This gives
\begin{equation} \label{Qmodel as tFS}
    f^{(n)}(x) = \sum_{\omega \in \Omega_n} c_\omega e^{i\omega x}
\end{equation}
where the coefficients $c_\omega$ are obtained by summing over all $a_{\bm{n}^{\prime \prime},\bm{n}^{\prime}}$ contributing to the same frequency
\begin{equation}
    c_\omega = \sum_{\substack{\bm{n}^{\prime \prime},\bm{n}^{\prime} \\ n^\prime_1 - n^{\prime \prime}_1 = \omega}} a_{\bm{n}^{\prime \prime},\bm{n}^{\prime}},
\end{equation}
with $c_\omega = c_{-\omega}^*$ and Eq.~\eqref{Qmodel as tFS} is a real-value function, as it should be. The frequency spectrum $\Omega_n = \{ n^\prime_1 - n^{\prime \prime}_1 \:| \: n^\prime_1,n^{\prime \prime}_1\in [0,n] \}$ contains all frequencies that are accessible to the $n$-photon quantum model. For general trainable circuit blocks $\mathcal{W}^{(i)}(\bm{\theta}_i)$, measurement observable $\mathcal{M}(\bm{\lambda})$ and $n$-photon Fock state $\ket{\bm{n}^{(i)}} = \ket{n^{(i)}_1,n^{(i)}_2,\dots,n^{(i)}_m}$, the $n$-photon quantum model reads
\begin{equation} \label{eqn:nphoton-Qmodel-1v}
    \begin{aligned}[b]
    f^{(n)}(x,\bm{\Theta},\bm{\lambda}) &= \sum_{\omega \in \Omega_n} c_{\omega}(\bm{\Theta},\bm{\lambda}) e^{i\omega x}.
    \end{aligned}
\end{equation}

\begin{figure}[t]
    \includegraphics[width=80mm,scale=0.5]{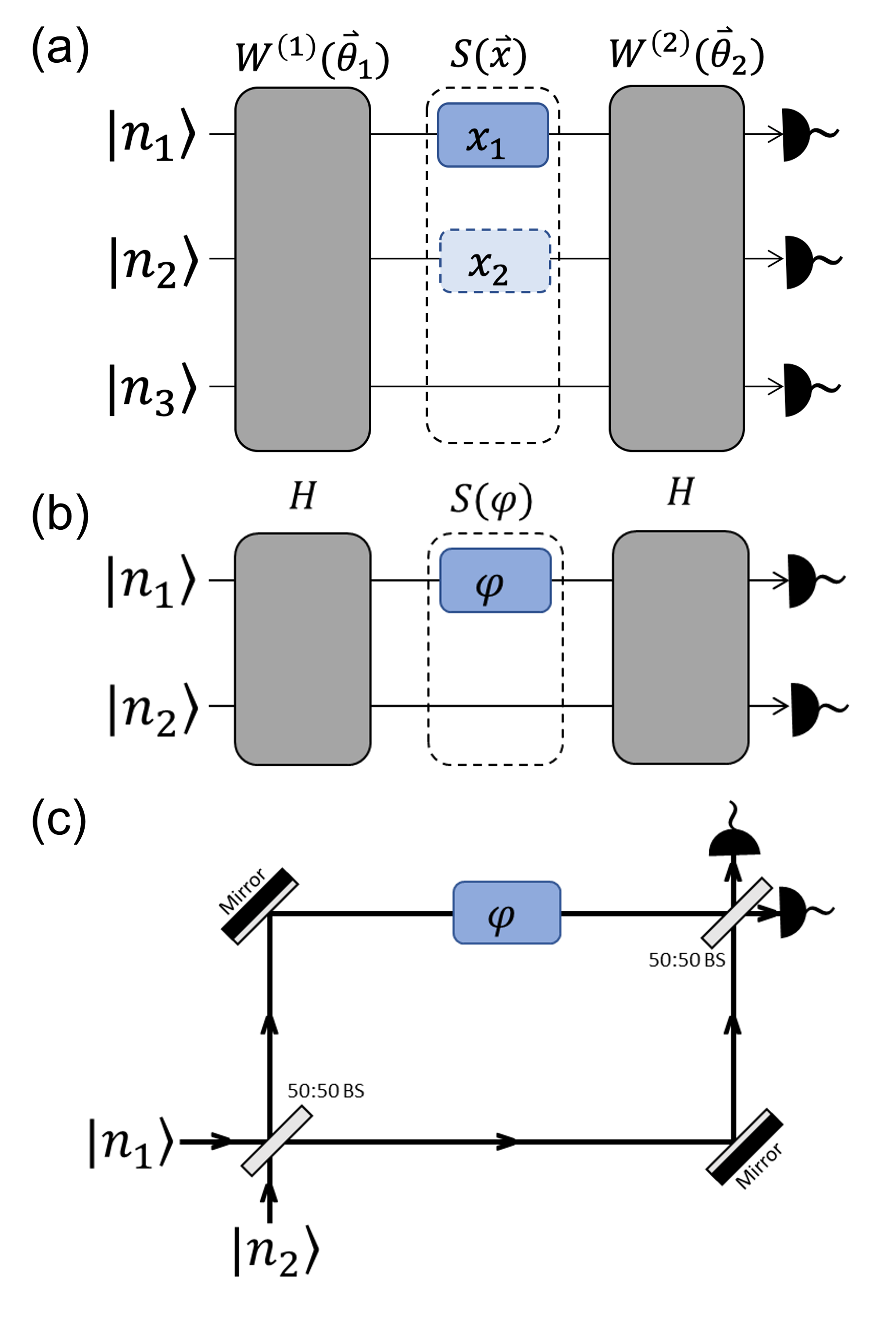}
    \caption{\label{fig:CircuitDiagram-3approaches} Different linear quantum photonic circuit configurations for supervised machine learning. (a) Parameterized circuit comprising three spatial modes for fitting of Fourier series and binary classification. One encoding phase shifter is used per classical data feature. (b,c) Two spatial mode circuits for implementing kernel-based machine learning using Gaussian kernels with photon number-resolving detectors. Here $H$ denotes a 50-50 beamplitter, with matrix elements the same as the Hadamard transform~\cite{motes2015linear,olson2017linear}. In other words, (b) is a (c) Mach-Zehnder interferometer. Direct implementation of the kernel method can be done by using the phase shifter to encodes the squared distance between pairs of data points, $\phi = \delta = ({\bf x}-{\bf x}')^2$, while random kitchen sink approach approximate a Gaussian kernel using a set of randomized input features $\phi = x_{r,i} = \gamma ( \bm{w}_r\cdot\bm{x}_i+ b_r)$.}
\end{figure}

\subsection{Expressive power and trainability of linear quantum photonic circuits}
Since the $n$-photon quantum model can be represented by a Fourier series, its expressive power can be studied via two properties: its frequency spectrum and Fourier coefficients. The former tells us which functions the quantum model has access to, while the latter determines how the accessible functions can be combined \cite{schuld2021effect}.

\subsubsection{\label{subsubsec:n-dependent-expressivity} Photon-number dependent frequency spectrum}
The frequency spectrum can be easily shown to be
\begin{equation} \label{nphoton-Frequency_spectrum}
    \Omega_n = \{-n,-n+1,\dots,n-1,n\},
\end{equation}
which is solely determined by the number of photons fed into the circuit. It always contains the zero frequency, i.e. $0 \in \Omega_n$, while the non-zero frequencies occur in pairs, i.e. $\omega, -\omega \in \Omega_n$. This motivates us to define the size of the frequency spectrum as $D_n = (|\Omega_n| -1)/2 = n$ to quantify the number of independent non-zero frequencies the $n$-photon quantum model has access to. In comparison to Ref.~\cite{schuld2021effect}, where the size of frequency spectrum is determined by the spectrum of the data encoding Hamiltonian, here the size of the frequency spectrum can be increased by feeding more photons into the circuit, while keeping the number of spatial modes and encoding phase shifters constant. 

This implies that $n$-photon quantum models with more input photons can be more expressive, because they have access to more basis functions, and hence can learn Fourier series with higher frequencies. In the limit of $n \rightarrow \infty$, i.e. continuous variable quantum systems, $n$-photon quantum models can support the frequency spectrum $\Omega_\infty = \{ -\infty,\dots,-1,0,1,\dots,\infty \}$ of a full Fourier series, in agreement with Ref.~\cite{schuld2021effect}. For a fixed number of input photons, the frequency spectrum can be broadened further using multiple encoding phase shifters, either in parallel or sequentially~\cite{schuld2021effect,perez2020data} (see Appendix~\ref{app:Encoding_scheme_1DFS} for details).

As an example, we consider training a linear QPC with three spatial modes shown in Fig.~\ref{fig:CircuitDiagram-3approaches}(a) using a regularized squared loss cost function. The cost function $C(\bm{\Theta},\bm{\lambda})$ is constructed using the measurement results and a training set of $N$ desired input/output pairs $\{x_i \rightarrow g(x_i) \}_{i=1}^N$
\begin{equation} \label{eqn: cost_func-univariable}
    C(\bm{\Theta},\bm{\lambda}) = \frac{1}{2N} \sum_{i=1}^N (g(x_i)-f^{(n)}(x_i,\bm{\Theta},\bm{\lambda}))^2 + \alpha \bm{\lambda} \cdot \bm{\lambda},
\end{equation}
that is variationally minimized over $\bm{\Theta}$ and $\bm{\lambda}$ to learn the function $g(x)$. Here, $f(x,\bm{\Theta},\bm{\lambda})$ is the $n$-photon quantum model in Eq.~\eqref{eqn:nphoton-Qmodel-1v}, while $\bm{\lambda} \cdot \bm{\lambda} = \sum_i \lambda_i^2$ is the sum of squared observable parameters, forming a regularization term with weight $\alpha$. The regularization term has a two-fold role: it prevents model over-fits, and ensures that the model prediction is not based on output photon combinations that occur with very low probability. The latter is important for QPC-based machine learning models, because the number of measurements required to obtain all of the required expectation values scales with the number of spatial modes and photons. 

\begin{figure}[t]
\includegraphics[width=80mm,scale=0.5]{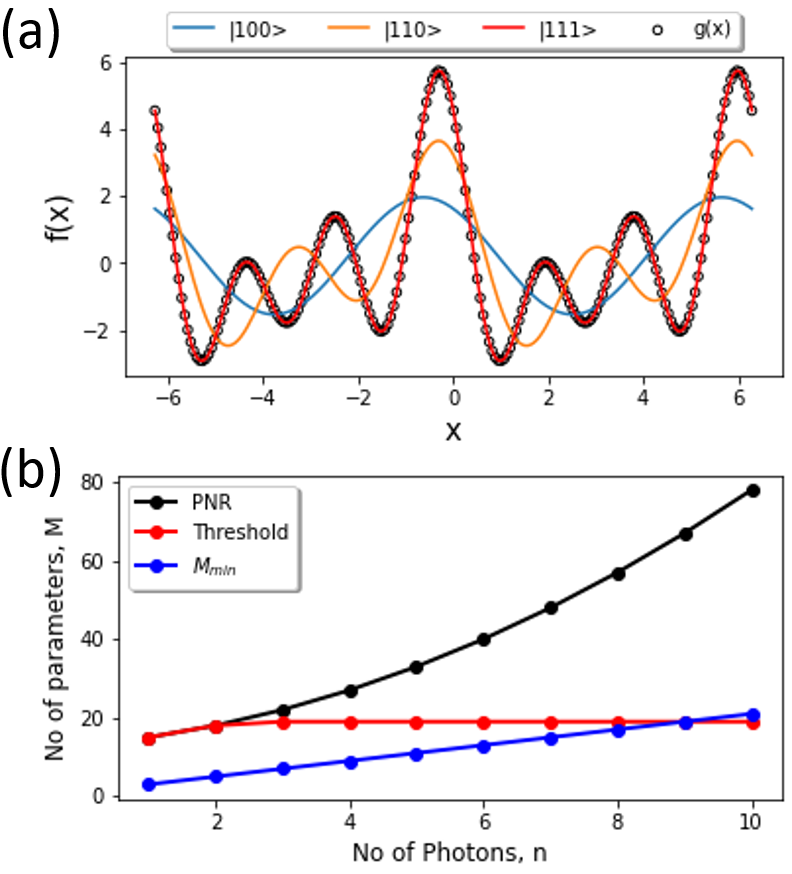}
\caption{\label{Schematic-Fitting_example-DOF} Photonic number-dependent expressive power of the variational linear quantum photonic circuit. (a) Optimal fits of a degree three Fourier series $g(x) = \sum_{n=-3}^{3} c_n e^{-nix}$ with coefficients $c_0 = 0.2$, $c_1 = 0.69+0.52i$, $c_2 = 0.81+0.41i$ and $c_3 = 0.68+0.82i$ using a three mode circuit with photon number-resolving (PNR) detectors and different input Fock states, i.e. $\ket{100}$, $\ket{110}$, and $\ket{111}$. A perfect fit is achieved using at least three input photons, since a sufficient number of non-zero frequencies are encoded, i.e. $D_n = 3$. (b) The number of real degrees of freedom of the three mode parameterized quantum photonic circuit with PNR detectors (black) and threshold detectors (red). The former is always larger than the minimum number of parameters $M_{\mathrm{min}}$ required to control all the circuit's Fourier coefficients (blue), and hence, arbitrary Fourier coefficients can be realized by this circuit. In contrast, the expressive power of the threshold detectors (red) can only be enhanced for up to 9 input photons.}
\end{figure}

We train the three mode linear QPC using the gradient-free algorithm in the \textit{NLopt} nonlinear-optimization python package \cite{johnson2014nlopt}, i.e. BOBYQA algorithm \cite{powell2009bobyqa} to fit a degree three Fourier series $g(x)$ with $x \in [-3 \pi, 3 \pi]$ using input states of up to 3 photons. We consider input states for which each spatial mode contains at most one photon, i.e. $\ket{\bm{n}^{(i)}} = \ket{100}$, $\ket{110}$, or $\ket{111}$. Fig.~\ref{Schematic-Fitting_example-DOF}(a) shows how the number of observable frequency components and hence the expressive power of the circuit grows with the number of input photons. Perfect fitting is achieved with three photons. In contrast, the frequency spectrum of similar qubit-based architectures cannot fit a degree two Fourier series using a single encoding block, requiring either a deeper or wider circuits with multiple encoding gates~\cite{perez2020data,schuld2021effect}. 

\subsubsection{\label{subsubsec: exp FC} Trainability of Fourier coefficients}

Even if the parametrized QPC can generate the frequency spectrum required to fit the desired function, this does not necessarily imply that the optimal Fourier coefficients are accessible~\cite{schuld2021effect}; the linear circuits we consider cannot perform arbitrary Fock state transformations. However, we do not need to generate arbitrary Fock states and only require control over one real and $D_n$ complex Fourier coefficients $\{c_{\omega} \}$. For $n$ input photons and taking $D_n = n$, this requires at least $M_{min}= 2D_n+1$ real degrees of freedom.

Each trainable circuit blocks has $m(m-1)$ controllable parameters \cite{reck1994experimental}, while the number of controllable degrees of freedom of the parameterized observable depends on type of detector. For photon number resolving (PNR) detectors the number of degrees of freedom is
\begin{equation} \label{eqn: DoF-PNR}
 M_{\text{PNR}} = 2m(m-1) + \frac{(n+m-1)!}{n!(m-1)!},
\end{equation}
while threshold detectors have
\begin{equation}
 M_{\text{THR}} = 2m(m-1) + \sum_{k=1}^{\min(n,m)}\frac{m!}{k!(k-m)!}
\end{equation}
degrees of freedom.

For a fixed number of spatial modes and photons, threshold detectors have fewer controllable degrees of freedom compared to PNR detectors, and hence their expressive power saturates beyond a certain number of input photons. For example, Fig.~\ref{Schematic-Fitting_example-DOF}(b) illustrates the expressive power of a circuit with three spatial modes. Using threshold detectors the expressive power is only enhanced by increasing the number of photons up to nine; beyond this, the number of controllable degrees of freedom is less than $M_{\min}$. On the other hand, using PNR detectors the circuit may in principle be trained to fit arbitrarily large frequencies by increasing the number of input photons. Of course, in practice the range of frequencies accessible using a single encoding gate will be limited by sensitivity to losses, which grows exponentially with the photon number.

\subsubsection{Universality of the linear quantum photonic circuit}

It is well known that a Fourier series can approximate any square-integrable function $g(x)$ on a finite interval to arbitrary precision \cite{carleson1966convergence,weisz2012summability}. Thus, expressing the $n$-photon quantum model in term of a Fourier series allows us to demonstrate the universality of the quantum model by studying its ability to realise arbitrary Fourier series. Universality of a Fourier series is determined by two important ingredients: a sufficiently-broad frequency spectrum and arbitrary realizable Fourier coefficients. The analysis in Sec.~\ref{subsubsec:n-dependent-expressivity} implies that the frequency spectrum $\Omega_n$ accessible by $n$-photon quantum models asymptotically contains any integer frequency if enough input photons are used, satisfying one of the criteria to achieve universality. 

To realize arbitrary set of Fourier coefficients, at least $M \ge M_{\min} = 2n+1 $ degrees of freedom in the linear QPC are required. Here, we consider a linear QPC with PNR detectors. The PNR detectors are used because the expressive power of threshold detectors saturated beyond some threshold number of photons. One of the unique advantages of photonic system is the exponentially growing dimension of the Fock space with number of spatial modes and photons. For a linear QPC with constant number of spatial modes $m$, the dimension of the Fock space and $M_{\text{PNR}}$ scales in the order of $O(n^{m-1})$, hence contributing $O(n^{m-1})$ of degrees of freedom. On the other hand, the degrees of freedom from the trainable circuit blocks scale with $O(m^2)$, which is negligible when $n \gg m$. By exploiting this advantage, it can be seen that $M_{\text{PNR}}$ is always larger than $M_{\min}$ as the size of the frequency spectrum $D_n$ and $M_{\min}$ scale linearly with photon number, i.e. $O(n)$. This is a necessary condition for the $n$-photon quantum model being able to realize arbitrary set of Fourier coefficients, which in the examples we consider also seems to be sufficient. More rigorously, following the arguments in Ref.~\cite{schuld2021effect} a universal function approximator may be obtained by generalizing our circuits to arbitrary (entangled) input states and observables by incorporating nonlinear elements into the circuits \footnote{The nonlinear optical elements enables the realization of two qubit entangling gates, while arbitrary single qubit rotations can be realized using linear optics such as beam splitter and phase shifters. By considering the dual-rail photonic qubit, our circuit could perform universal quantum computation, and the Fourier coefficients arguments follows from Ref.\cite{schuld2021effect}.}.

As an example, we consider a linear QPC with 3 spatial modes. In this case, Eq.~\eqref{eqn: DoF-PNR} is reduced to
\begin{equation}
 M_{\text{PNR}} = 12 + \frac{(n+2)(n+1)}{2!},
\end{equation}
which is always larger than $M_{\min} = 2n+1$ for $n \in \mathbb{N}$, as shown in Fig.~\ref{Schematic-Fitting_example-DOF}(b). Hence, the $n$-photon quantum model with 3 spatial modes and a single phase shifter can act as a universal function approximator. In contrast, the qubit-type variational quantum circuits require deep or wide circuits and many encoding gates to ensure a rich frequency spectrum, and arbitrary global unitaries to realize arbitrary sets of Fourier coefficients~\cite{schuld2021effect}.  

\subsubsection{Effect of noise on the expressive power of linear quantum photonic circuits}
For noiseless linear quantum photonic circuits, we have shown that its expressive power will improve with the increasing number of photons and spatial modes. In this section, we will discuss the role of optical losses on the expressive power of linear quantum photonic ciruits. For real quantum photonic hardware, the optical loss sensitivity will grows exponentially with the circuit depth and number of input photons. The typical noise sources are (1) inefficient collection optics, (2) losses in the optical components due to absorption, scattering, or reflections from the surfaces, (3) inefficiency in the detection process due to using detectors with imperfect quantum efficiency, and they can be modelled using beam splitters \cite{fox2006quantum}. These noises will obviously affect the frequency spectrum, where the higher frequency term cannot be distinguished from the lower frequency term, hence reducing the size of the frequency spectrum. We anticipate the noises to have a minimum impact on the Fourier coefficients, as they depend only on the physical components such as linear optics in trainable blocks and the detectors. Therefore, the output observables can still be written as Fourier series, just with reduced expressivity. This will place a practical limit on the complexity of the QML models using this scheme, unless one can include some kind of error correction scheme. When the losses are low enough, the detectors should have a sufficiently high signal to noise ratio that other noise sources can be neglected. Apart from the error correction scheme, the regularization term in the cost function should be able to help to minimize the detrimental influence of noise. It penalizes models including coefficients with huge weights, hence no particular output state should have a huge weight, reducing the model's sensitivity of noise in final prediction. Finally, the photonic circuit considered here are based on variational approach, therefore, they are robust against variations in the beam splitter ratios, tuning, and etc. The quantitative noise modelling of the linear photonic quantum circuits will be a subject for future research.

\section{\label{sec:classification} Supervised Learning using linear quantum photonic circuits}

\begin{figure*}
\includegraphics[width=175mm,scale=0.5]{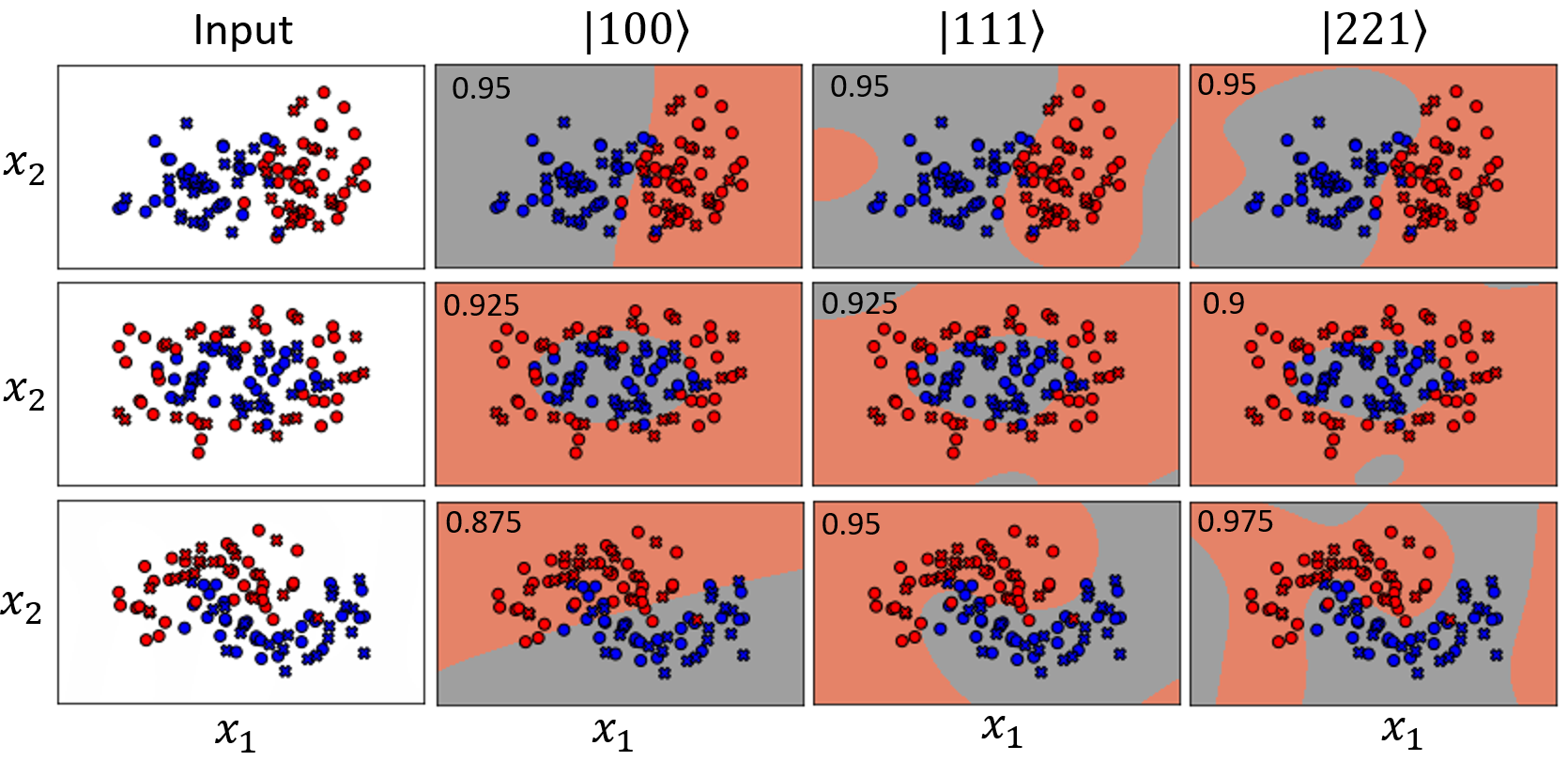}
\caption{\label{fig:Variational-classification-results} Binary classification using the three mode linear quantum photonic circuit of Fig.~\ref{fig:CircuitDiagram-3approaches}(a) with different input Fock states $\ket{100}, \ket{111}$, and $\ket{221}$, training using 60 points with a regularization weight $\alpha = 0.2$. First row: linearly-separable dataset. Middle row: circle dataset. Bottom row: moon dataset. The performance on a test set (red and blue solid cross) of 40 points is given in the upper left corner of each respective subplot. The classification boundaries for all datasets become more complicated as the number of input photons increases, illustrating the increasing expressive power. Increase of expressive power does not affect the trainability of the linear dataset, since a linear classifier suffices. The performance for the circle dataset degrades for larger input photons due to over-fitting, demonstrating that a larger expressive power is not necessarily better. On the other hand, a higher expressive power is necessary in order to accurately classify the moon dataset.}
\end{figure*}

As an application of the trainable linear QPCs we now consider different strategies for binary classification. In the first strategy the linear QPCs are directly used as variational quantum classifiers, classifying data directly on the high-dimensional Fock space by optimizing a regularized squared loss cost function. In this case, as the circuit becomes more expressive it becomes harder to train. Second, we consider kernel methods as a means of avoiding the costly circuit optimization step. We show how linear circuits can be used to implement Gaussian kernels either directly or using the random kitchen sinks algorithm, sampling kernels with different resolutions in parallel. Note that we are mainly interested in what kinds of kernel functions can be efficiently implemented using linear QPCs, instead of providing quantum kernels that might offer quantum advantages, motivated by ongoing interest in classical photonic circuits for machine learning~\cite{de2019photonic,hamerly2019large,roques2020heuristic,shastri2021photonics}.

\subsection{Linear quantum photonic circuit as variational quantum classifiers} \label{subsec:variational-classifier}

We perform binary classification of two-dimensional classical data. Each data dimension is encoded using a single phase shifter, as shown in Fig.~\ref{fig:CircuitDiagram-3approaches}(a). The mapping of the data into the high-dimensional Fock space is nonlinear, circumventing the need for nonlinear optical elements.

The $n$-photon supervised classification model for two-dimensional data $f^{(n)}(\bm{x},\bm{\Theta},\bm{\lambda})$ is defined as
\begin{equation}
    \begin{aligned}[b]
    f^{(n)}(\bm{x},\bm{\Theta},\bm{\lambda})= \sum_{\bm{\omega} \in \Omega_n} c_{\bm{\omega}}(\bm{\Theta},\bm{\lambda}) e^{i\bm{\omega} \cdot \bm{x}}
    \end{aligned}
\end{equation}
where $\bm{x} = (x_1,x_2)$ is the 2-dimensional data feature, $\bm{\omega} = (\omega_1, \omega_2)$ is the 2-dimensional frequency vector, and $\Omega_n = \{ -\bm{\omega},0,\bm{\omega}$ $|$ $\omega_1, \omega_2 \in [0,n]$, $\omega_1+\omega_2 \le n \}$ is the frequency spectrum of the model. This encoding scheme will not generate a full frequency spectrum for multi-dimensional Fourier series but it suffice for the example considered here. See Appendix~\ref{app:Encoding_scheme_multiDFS} for schemes that generate full frequency spectrum for multi-dimensional Fourier series.

The model is trained by minimizing the cost function
\begin{equation} \label{Eqn:squared loss multi-D}
    C(\bm{\Theta},\bm{\lambda}) = \frac{1}{2N} \sum_{i=1}^N (g(\bm{x}_i)-f^{(n)}(\bm{x}_i,\bm{\Theta},\bm{\lambda}))^2 + \alpha \bm{\lambda} \cdot \bm{\lambda}
\end{equation}
using the BOBYQA algorithm, with the decision boundary defined as 
\begin{align} \label{eqn:eqn-decision-boundary}
    f^{(n)}_{\text{sgn}}(\bm{x}) = \text{sgn}[f^{(n)}(\bm{x},\bm{\Theta}_{opt},\bm{\lambda}_{opt})]
\end{align}
where $\bm{\Theta}_{opt}$ and $\bm{\lambda}_{opt}$ are the optimized circuit's and observable's parameters and $\mathrm{sgn}$ is the sign function. Thus, the class of the data points is assigned by the sign of circuit output.

As an example, we trained the linear QPC to classify three different types of datasets from the \textit{scikit-learn} machine learning library \cite{scikit-learn}: linear, circle, and moon. Fig.~\ref{fig:Variational-classification-results} illustrates the trained models. The contour plots show that $n$-photon supervised classification models with higher photon number have more complicated classification boundaries, in agreement with previous analysis on the expressive power of quantum models. Since the linear data set can be separated by a linear decision boundary, unsurprisingly a single model photon is sufficient to learn the classification boundary. On the other hand, overfitting can occur when the model expressive power is too large, as can be seen for the degraded performance for the circle dataset for the $\ket{221}$ input state. The classification performance for the more complicated moon dataset improves with the number of input photons. These examples illustrate the impact of a higher expressive power on classification using linear QPCs. 

\subsection{Linear quantum photonic circuits as Gaussian kernel samplers}

\begin{figure}[t]
\includegraphics[width=80mm,scale=0.5]{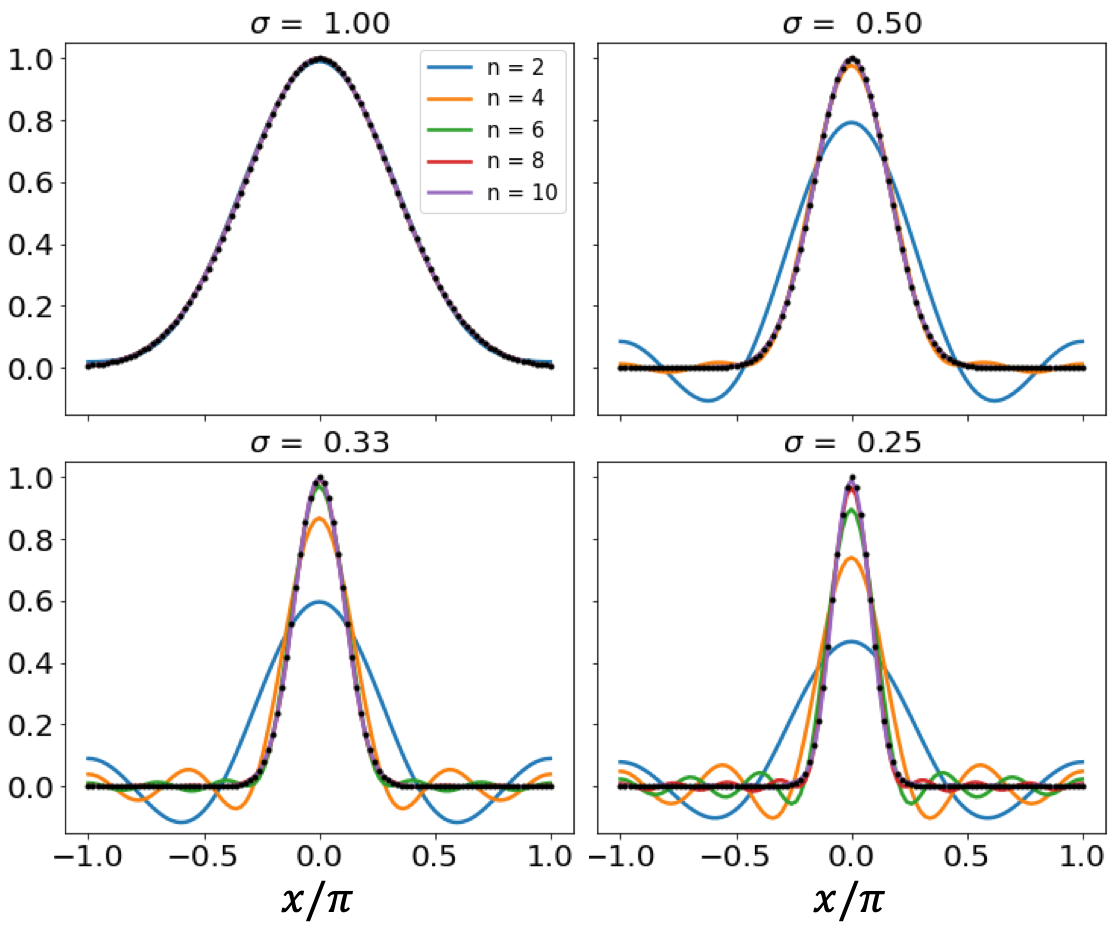}
\caption{\label{Gaussian_kernel-kernel_methods} Approximating Gaussian kernels with different resolutions $\sigma = 0.25,0.33,0.50,1.00$ using the two mode linear quantum photonic circuit of Fig.~\ref{fig:CircuitDiagram-3approaches}(b) with different numbers of input photons $n = 2, 4, 6, 8, 10$. Two photons are sufficient to approximate a low resolution kernel with $\sigma = 1.00$ (curves for all photon numbers overlap), while higher resolutions require more photons to approximate. For example, a circuit with four photons can fit Gaussian kernels with $\sigma = 0.50$, but not $\sigma = 0.33$ or $0.25$.}
\end{figure}

Similar to the standard noisy and large scale variational circuits, the variational machine learning approach becomes more difficult to train as the dimension of the Fock space increases, likely due to the issue of vanishing cost function gradients \cite{mcclean2018barren,wang2020noise,marrero2020entanglement,bittel2021training,thanasilp2021subtleties}, requiring exponentially-growing precision to optimize the circuit parameters in-situ~\cite{arrasmith2020effect}.
In addition, it is expensive to train the quantum gates (in this case the tunable beam splitter meshes) in the noisy-intermediate scale quantum (NISQ) era as it is time-consuming to reprogram quantum circuits~\cite{harrow2020small}. Due to these limitations, it is more efficient to use NISQ devices as sub-routines for machine learning algorithms, e.g. to sample quantities that are useful for classical machine learning models but time-consuming to compute. In particular, variational quantum circuits can be used to approximate kernel functions for classical kernel models such as support vector machines~\cite{bartkiewicz2020experimental,schuld2019quantum,havlivcek2019supervised,goto2021universal,haug2021large}. Here, we show how the linear QPCs can be designed to approximate Gaussian kernels with a range of resolutions determined by the number of input photons. 

\subsubsection{Kernel methods}
Kernel methods allow one to apply linear classification algorithms to datasets with nonlinear decision boundaries \cite{scholkopf2002learning,hofmann2008kernel}. The idea is to leverage feature maps $\phi(\bm{x})$ that map the nonlinear dataset from its original space into a higher dimensional feauture space in which a linear decision boundary can be found, enabling classification via a linear regression
\begin{equation*}
    f(\bm{x}) = \bm{w} \cdot \phi (\bm{x}),
\end{equation*}
using suitably-trained weights $\bm{w}$. Instead of computing and storing the high-dimensional feature vector $\phi$, the kernel trick \cite{mercer1909functions,hofmann2008kernel} is employed by introducing a kernel function $k(\bm{x},\bm{x}')$, which measures the pairwise similarity between the data points in the feature space. Formally, the kernel functions is defined as the inner product of two feature vectors 
\begin{equation}
    k(\bm{x},\bm{x}') = \phi(\bm{x}) \cdot \phi(\bm{x}').
\end{equation}
According to representer theorem \cite{scholkopf2001generalized}, the solution to the decision boundary can then be expressed in term of the kernel functions as
\begin{equation} \label{eqn:f-kernel methods}
    f(\bm{x}) = \sum_{i=1}^N\beta_i k(\bm{x}_i,\bm{x}),
\end{equation}
converting the optimization problem into a convex optimization problem of finding the parameters $\beta_i$. In the case of a regularized squared loss cost function such as Eq.~\eqref{Eqn:squared loss multi-D} the optimal parameters $\beta_i$ can be obtained analytically \cite{bishop2006pattern,theodoridis2015machine} as 
\begin{equation}
    \bm{\beta} = (K + \alpha \bm{I})^{-1}\bm{y}.
\end{equation}
where $N$ is the number of training data, $K$ is the $N \times N$ kernel matrix with matrix elements $K_{ij} = k(\bm{x}_i,\bm{x}_j)$, $I$ is the $N$-dimensional identity matrix, $\alpha$ is the regularization parameter, and $\bm{y}$ is the $N \times 1$ vector of the training data labels. 

Although we currently have an example that shows rigorous performance guarantees of quantum kernel methods on artificial dataset~\cite{liu2021rigorous}, it is still unclear whether quantum machine learning models can achieve improved performance compared to classical machine learning approaches in practical problems by sampling from kernels that are hard to compute classically ~\cite{schuld2019quantum,schuld2021machine,huang2021power,wang2021towards}. Even in the absence of a rigorous quantum advantage, special-purpose electronic and photonic machine learning circuits are being pursued in order to increase the speed and energy-efficiency of well-established classical machine learning models~\cite{de2019photonic,hamerly2019large,roques2020heuristic,shastri2021photonics}. Therefore, here we will focus on implementing the widely-used Gaussian kernel
\begin{equation*}
    k(\bm{x},\bm{x}') = k(\bm{x}-\bm{x}') = e^{-\frac{1}{2\sigma^2}(\bm{x}-\bm{x}')^2}
\end{equation*}
with controllable resolution $\sigma$. The Gaussian kernel is a universal, infinite-dimensional kernel that can learn any continuous function in a compact space~\cite{micchelli2006universal}.

\subsubsection{Linear quantum photonic circuits as sub-routine of kernel methods} \label{subsubsec: LQPC as sub-routine of KM}
We approximate the Gaussian kernel using the two mode QPC shown in Fig.~\ref{fig:CircuitDiagram-3approaches}(b), where 50-50 beamsplitters $\mathcal{H}$ are used for both trainable circuit blocks and the squared Euclidean distance between pairs of data points $\delta = (\bm{x}-\bm{x}^{\prime})^2$ is encoded using a single phase shifter. The output of this circuit can be written as

\begin{align*}
    f^{(n)}(\delta,\bm{\lambda}) &= \bra{n,0}\mathcal{U}^\dagger(\delta) \mathcal{M}(\bm{\lambda}) \mathcal{U}(\delta) \ket{n,0}\\
    &= \sum_{ij} \lambda_{ij}|\bra{n_i,n_j}\mathcal{U}(\delta)\ket{n,0}|^2\\
    &= c_0^{(n)}(\bm{\lambda}) + 2\sum_{k=1}^n c_k^{(N)}(\bm{\lambda})\cos(k\delta).
\end{align*}
where $\mathcal{U}(\delta) = \mathcal{H} \mathcal{S}(\delta) \mathcal{H}$ and the trainable observable $\mathcal{M}(\bm{\lambda}) = \sum_{ij} \lambda_{ij}\ket{n_i,n_j}\bra{n_i,n_j}$ with $n_i + n_j = n$. Similar to Sec.~\ref{subsec:variational-classifier}, the observable is trained to approximate the Gaussian kernel of resolution $\sigma$ by minimizing the squared loss cost function using the BOBYQA algorithm 
\begin{align*}
    f^{(n)}(\delta, \bm{\lambda}^{(\sigma)}) \approx e^{-\frac{\delta}{2\sigma^2}}.
\end{align*}
This approach has two advantages: Different kernel resolutions can be accessed using the same photon detection statistics $|\bra{n_i,n_j}\mathcal{U}(\delta)\ket{n,0}|^2$ by taking different linear combinations of the output observables $\mathcal{M}(\bm{\lambda}^{(\sigma)})$. Second, this training only needs to be performed once; the tunable circuit blocks do not need to be reconfigured if the training data set changes.

We note that the domain of the input data, in this case the norm squared distances between pairs of data points, must lie within the interval that defines the circuit's Fourier series. This imposes an upper bound on the kernel resolution that the linear QPC has access to. The circuit with higher expressive power, i.e. higher number of input photons, can more precisely approximate kernels with higher resolution $\sigma$. Kernels with lower resolution can already be well-approximated by a circuit with only two input photons. Fig.~\ref{Gaussian_kernel-kernel_methods} shows the kernel training result for different desired resolutions and input photon numbers. Once the kernel has been trained, classification can be performed by feeding the measured similarity matrix into a classical machine learning model such as a support vector machine \cite{steinwart2008support}. 

\subsection{Quantum-enhanced random kitchen sinks}

\begin{figure*}
\includegraphics[width=175mm,scale=0.5]{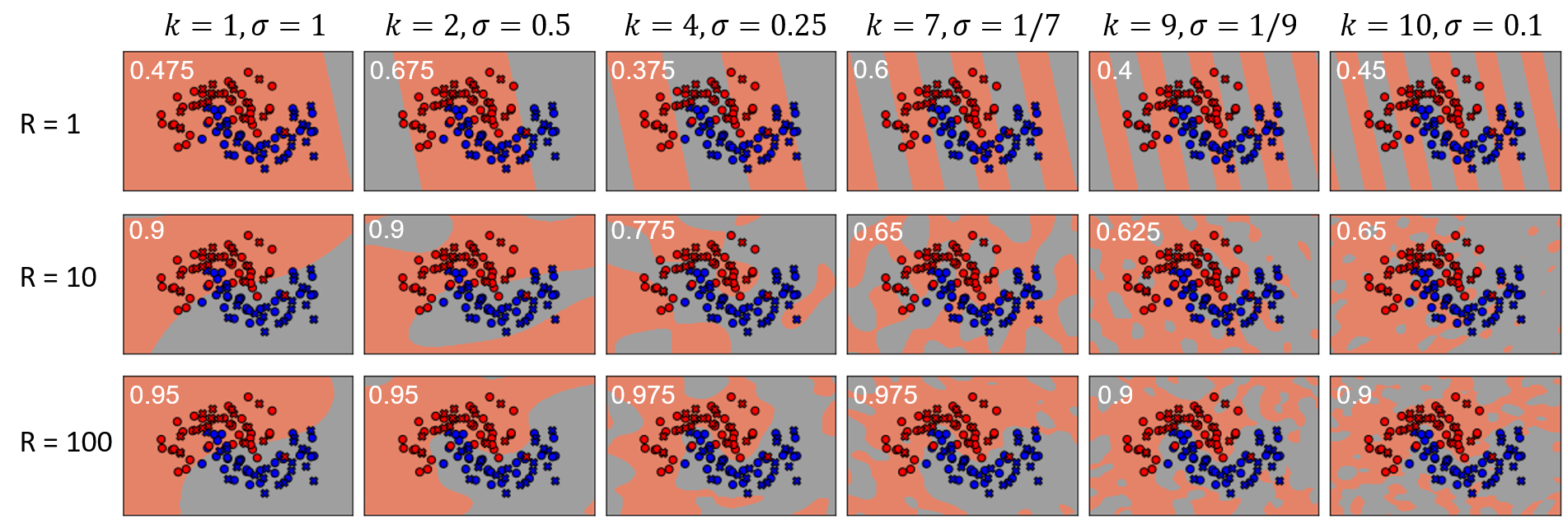}
\caption{\label{fig:classification-results-rff} Binary classification of the moon dataset of Fig.~\ref{fig:Variational-classification-results} using the two mode linear quantum photonic circuit of Fig.~\ref{fig:CircuitDiagram-3approaches}(c) which implements a quantum-enhanced random kitchen sink with 10 input photons, base (single input photon) resolution $\gamma = 1$, regularization parameter $\alpha = 0.2$, and random Fourier feature dimensions $R = 1, 10, 100$. The circuit with 10 input photons can probe 10 different kernel resolutions simultaneously, i.e. $\sigma = \{1/n \; | \; 1 \leq n \leq 10\}$; six resolutions are illustrated here. When $R = 1$ the feature vectors reduce to a cosine-like kernel whose frequency increases with the number of input photons and $k$. The classification results improve with $R$ because the kernels are better approximated by random Fourier features of higher dimension with $\sigma = 0.25$ and $1/7$ ($R = 100$), which are the optimal resolutions for the moon dataset. For a given $R$, the decision boundaries for higher resolutions are noisier because the corresponding approximated kernel has a narrower peak, thus meaningful predictions cannot be made for points that are far (relative to the kernel resolution) from the training set.}
\end{figure*}

One limitation of kernel methods is their poor (quadratic) scaling with the size of the training data set. To circumvent this issue, the random kitchen sinks (RKS) algorithm was developed, which uses randomly-sampled feature maps in order to controllably approximate desired kernels and more efficiently train classical machine learning models~\cite{rahimi2007random,rahimi2008uniform,rahimi2008weighted}. In particular, sampling from random Fourier features enables approximation of the Gaussian kernel. This motivates us to propose a quantum-enhanced RKS algorithm, where the subroutine of RKS algorithm, i.e. the random feature sampler is replaced by linear QPCs. The linear QPCs can simultaneously sample random Fourier features of different frequencies, providing a unique advantage compare to the qubit-based architecture. Our approach differs from previous proposals for quantum random kitchen sinks by directly constructing the kernel functions using the random Fourier features, instead of performing linear regression with random feature bit strings sampled from variational quantum circuits~\cite{wilson2018quantum,noori2020analog}.

\subsubsection{Random kitchen sinks}

The randomized $R$-dimensional vectors known as the random Fourier features are defined as
\begin{equation} \label{eqn:random-Fourier-feature}
	\bm{z}(\bm{x}) = 
	\frac{1}{\sqrt{R}} \begin{pmatrix}
		z_{\bm{w}_1}(\bm{x}) \\ z_{\bm{w}_2}(\bm{x}) \\ \dots \\ z_{\bm{w}_R}(\bm{x})
	\end{pmatrix}
\end{equation}
where each $z_{\bm{w}_r}(\bm{x})$ is a randomized cosine function 
\begin{align} \label{Eqn:randomized-cosine}
	z_{\bm{w}_r}(\bm{x}) = \sqrt{2} \cos(\gamma[\bm{w}_r \cdot \bm{x}+b_r]),
\end{align}
$\bm{x}$ is the $D$-dimensional input data, $\bm{w}_r$ are $D$-dimensional random vector sampled from a spherical Gaussian distribution, and $b_{r}$ are random scalars sampled from a uniform distribution,
\begin{align*}
	\bm{w} \sim \mathcal{N}_D(0,\bm{I}); \quad b \sim \text{Uniform}(0,2\pi).
\end{align*}
The random Fourier features approximate the Gaussian kernel~\cite{rahimi2007random,rahimi2008uniform}
\begin{equation} \label{eqn:approximating kernel with RFF}
	\bm{z}(\bm{x}) \cdot \bm{z}(\bm{x}') \approx k(\bm{x},\bm{x}') = e^{-\frac{\gamma^2}{2}(\bm{x}-\bm{x}')^2}
\end{equation}
with $\gamma$ acting as a hyperparameter that controls the kernel resolution. Note that other commonly-used kernels including Laplacian and Cauchy kernels can be approximated using the RKS algorithm using different sampling distributions~\cite{rahimi2007random}. 

Substituting Eq.~\eqref{eqn:approximating kernel with RFF} into Eq.~\eqref{eqn:f-kernel methods} yields
\begin{align} \label{eqn:f-rFF}
    f(\bm{x}) & \approx  \sum_{i=1}^N\beta_i \bm{z}(\bm{x}_i) \cdot \bm{z}(\bm{x}) = \bm{c} \cdot \bm{z}(\bm{x}).
\end{align}
where $\bm{c} = \sum_{i}\beta_i \bm{z}(\bm{x}_i)$. The optimal solution for a supervised learning problem using training data $\{(\bm{x}_j, y_j) \}_{j=1}^{N}$ that minimizes the regularized squared loss cost function \cite{hoerl1970ridge,bishop2006pattern,theodoridis2015machine} in Eq.~\eqref{eqn:f-rFF} is
\begin{align} \label{eqn:optf-rFF}
    f^*(\bm{x}) = \bm{c}_{\text{opt}} \cdot \bm{z}(\bm{x}).
\end{align}
where $\bm{c}_{\text{opt}}  = (\bm{z}(\bm{X})^T \bm{z}(\bm{X}) + \alpha I_R )^{-1}\bm{z}(\bm{X})^T\bm{y}$,  $\bm{z}(\bm{X})$ is an $N \times R$ matrix of the training data
\begin{align}
    \bm{z}(\bm{X}) = 
    \begin{pmatrix}
		\bm{z}(\bm{x}_1)^T \\ \bm{z}(\bm{x}_2)^T \\ 
		\vdots \\
		\bm{z}(\bm{x}_N)^T
	\end{pmatrix},
\end{align}
and $I_R$ is the $R$-dimensional identity matrix.

The RKS algorithm addresses the poor scaling of kernel methods with the number of data points by mapping the data into a randomized low-dimensional feature space, turning Eq.~\eqref{eqn:f-kernel methods} into a linear model on the $R$-dimensional vectors $\bm{z}(\bm{x})$. The complexity of finding the analytical solution for the coefficients is reduced from $O(N^3)$ to $O(R^3)$, saving enormous amounts of resources when $R \ll N$ while maintaining model performance comparable to standard classification methods~\cite{rahimi2007random,rahimi2008uniform}. 

\subsubsection{Linear quantum photonic circuits as random Fourier feature samplers}

Using the same circuit as in Sec.~\ref{subsubsec: LQPC as sub-routine of KM} with a randomized input encoding [Fig.~\ref{fig:CircuitDiagram-3approaches}(b)], i.e. $x_{r,i} = \gamma (\bm{w}_r \cdot \bm{x}_i+ b_r)$, the circuit output becomes
\begin{align}
    f^{(n)}(x_{r,i}, \bm{\lambda}) = c_0^{(n)}(\bm{\lambda}) + 2\sum_{k=1}^n c_k^{(n)}(\bm{\lambda})\cos(kx_{r,i}).
\end{align}
Constructing different observables $\mathcal{M}(\bm{\lambda}^{(k)})$ from the same photon detection statistics $|\bra{n_i,n_j}\mathcal{U}(x_{r,i})\ket{n,0}|^2$ allows one to isolate cosine functions with different frequencies $k$ 
\begin{align} \label{Eqn:randomized-cosine-Q}
    f^{(n)}(x_{r,i}, \bm{\lambda}^{(k)}) = \sqrt{2}\cos(k\gamma [\bm{w}_r\cdot \bm{x}_i+b_r]),
\end{align}
which has the same structure as Eq.~\eqref{Eqn:randomized-cosine} with $\gamma \rightarrow k \gamma$. Thus, constructing the random Fourier features in Eq.~\eqref{eqn:random-Fourier-feature} with the randomized cosine functions Eq.~\eqref{Eqn:randomized-cosine-Q} enables us to approximate the kernel
\begin{equation}
	k(\bm{x},\bm{x}') = e^{-\frac{k^2 \gamma^2}{2}(\bm{x}-\bm{x}')^2}
\end{equation}
with resolution $\sigma = \frac{1}{k \gamma}$. In other words, Gaussian kernels with different resolutions can be accessed using a single QPC and the same set of measurements by considering different observables. The number of kernel resolutions accessible by the circuit is equal to the size of the frequency spectrum, i.e. circuits with more input photons have access to more resolutions. Here, the photon-number dependent expressive power of the linear QPCs is leveraged to produce a linear combination of cosine functions of different frequencies, simultaneously producing multiple random Fourier features that approximate Gaussian kernels of different resolutions.

Fig.~\ref{fig:classification-results-rff} illustrates the performance of moon dataset classifiers using circuits with 10 input photons and random Fourier features of different dimensions, i.e. $R = 1, 10, 100$ and the same decision boundary as in Eq.~\eqref{eqn:eqn-decision-boundary}. The circuit with input 10 photons can probe a range of kernel resolutions within one order of magnitude, e.g: for $\gamma=1$, the accessible resolutions are $\sigma =  \{1/n \; | \; 1 \leq n \leq 10\}$; six of these are shown in Fig.~\ref{fig:classification-results-rff} to illustrate the working principle of quantum-enhanced RKS. The decision boundary of smaller $\sigma$ is considerably noisier than for larger $\sigma$. This is because the kernel with smaller resolution has a narrower peak, and hence, predictions far away from the training data points cannot be made. The random Fourier features with higher dimensionality provides a better approximation to the kernel, thus suppressing the noise around training data points while improving the classification accuracy. The optimal resolution for the moon dataset is $\sigma = 0.25$ and $1/7$ for $R = 100$.

\subsection{Resource requirements for each scheme}

Each of the classification methods has different strengths and limitations in terms of the resource requirements, i.e. number of distinct circuit evaluations for performing training and predictions, summarized in Tab.~\ref{tab:Resource comparison}. Here, we are concerned only with number of distinct circuit evaluations required to perform training and prediction, since the quantum resources are much more precious than the classical resources in the NISQ era. For the variational circuit, the data features are directly encoded, but the beam splitters and phase shifters in the trainable circuit blocks need to be optimized in the training step. Hence, the training resource per optimization loop is $O(NDM)$, where $N$, $D$, and $M$ are the number of training data, the dimension of the data features, and the number of trainable circuit and observable parameters, respectively. More Fourier frequencies can be obtained with larger photon numbers but require larger $M$ for universality, increasing the training time. Prediction requires reconfiguration of the $D$ encoding phase shifters, which can however be performed in parallel.

\begin{table} 
\caption{\label{tab:Resource comparison} \textit{Quantum resource requirements for different schemes.}  The resource requirements for performing training and prediction are defined in term of the number of linear optical element (i.e. beam splitters and phase shifters) settings required, where $N$ is the number of training data, $D$ is the dimension of the data features, $M$ is the number of trainable circuit and observable parameters, and $R$ is the number of random Fourier features.}
    \begin{ruledtabular}
        \begin{tabular}{lcc}
        \textrm{Schemes} & \textrm{Training time} & Prediction time\\
        \colrule
        Variational methods & $O(NDM)$ & D\\
        Kernel methods & $O(N^2)$ & N\\
        Random kitchen sinks & $O(NR)$ & R\\
        \end{tabular}
    \end{ruledtabular}
\end{table}

Kernel methods, on the other hand, encode the differences between data inputs using one phase shifter and the training is outsourced to a classical computer, therefore the resources for training scale only with the number of training data, i.e. $O(N^2)$. The Gaussian kernel with different resolutions can be accessed with a fixed circuit by considering different observables; Gaussian kernels with higher resolutions are better approximated for circuits with larger numbers of input photons. In contrast to the variational methods, $N$ different phase shifter settings are required to make predictions on new data. Random kitchen sinks have similar advantages to kernel methods, i.e. fixed circuit and different resolutions can be accessed by different observables, but have a better scaling $O(NR)$ with number of input data points, where $R$ is the number of random features chosen. The predictions require $R$ circuit settings regardless of the dimension of the data features.

\section{Conclusion}
\label{sec:conclusion}

The data-embedding process is a bottleneck which must be addressed~\cite{harrow2020small} in order to fully leverage the potential of quantum machine learning algorithms. In this paper, we addressed the data encoding problem by proposing a more gate-efficient bosonic encoding method. Our method has three potential advantages. First, it allows for a more efficient data encoding by modulating all Fock basis simultaneously using only one phase shifter, regardless of the input photon number. Second, the circuits employed a kernel-like trick, where nonlinearity is outsourced to quantum feature maps, i.e. the data-encoding phase shifter that encode the classical data into the high-dimensional Fock space~\cite{havlivcek2019supervised,schuld2019quantum}, avoiding the need of the experimental hard-to-implement nonlinear optical components. Subsequently, the expressive power of the circuit can be controlled by the number of input photons, while requiring fewer encoding layers compared to the qubit-based architecture \cite{schuld2021effect,perez2020data}. Finally, the circuits can be trained to implement commonly-used kernels with well-understood properties such as the Gaussian kernel.

Even though our photonic models are inspired by the BosonSampling circuits \cite{aaronson2011computational}, we do not expect the arguments about the BosonSampling's classical non-simulability to hold for our circuits, for three reasons: (1) The model output is expectation values, not samples. (2) Our phoronic circuits are not sampled from the Haar random distribution. (3) The assumption of $m = n^2$ is relaxed, where $m$ is the number of optical modes and $n$ is the number of input photons. Even so, there exist other benefits of studying the use of this class of circuit as quantum machine learning models. Quantum machine learning is still in its infancy, and it is still unclear how to rigorously define a quantum advantage for generic machine learning problems \cite{schuld2022quantum}. In this work, we focused on a specific problem in this field, the data-encoding problem, showing using simple quantum machine learning models how bosonic circuits may enable more efficient data uploading. We expect our conclusions to be valid for other classes of quantum machine learning models which may be hard to classically simulate. In addition, we believe our photonic models will serve as primitive quantum machine learning model \cite{schuld2022quantum} that inspire researchers in the field to develop other photonic quantum machine learning algorithms that possess quantum advantages. Recently, the awareness of the importance of the energy aspect of quantum algorithms has been raised \cite{auffeves2021quantum}. Although the energy aspect of our quantum circuits is not studied, our models could inspire an application-oriented framework to compare the energy consumption of quantum machine learning based on different platforms.

While the dimension of the Fock space grows (exponentially) with photon number and spatial modes, improving the expressive power, this is accompanied by higher sensitivity to optical losses and the need for more detection events in order to accurately sample all of the required output observables. Moreover, there exists a trade-off between the model's expressive power and ability to generalize, where circuits with higher expressive power can suffer from larger generalization errors, i.e. over-fitting~\cite{banchi2021generalization,caro2021encoding} and trainability issue \cite{holmes2022connecting}. One potential way to mitigates these issues is to define the quantum machine learning models in projected Fock spaces, which may lead to potential quantum advantages \cite{huang2021power}.

We proposed three different ways with different resource requirements to perform binary classification using linear quantum photonics circuits (QPCs). (1) Variational quantum classifiers that classify data points directly on the high-dimensional Fock space, while (2) and (3) implement Gaussian kernel for classical kernel machines directly or using the random kitchen sinks algorithm, sampling kernels with different resolutions in parallel. The random kitchen sink approach could be further improved by sampling the random features from a data-optimized distribution using fault-tolerant quantum computers~\cite{yamasaki2020learning,Yamasaki2021ExponentialEC}. A linear QPC with three spatial modes and up to 10 input photons equipped with photon-number resolving detectors is sufficient to show a proof of concept experiment for all of the proposed approaches. Therefore, our proposed architecture can be implemented with current technology, such as integrated photonic circuits \cite{carolan2015universal,Zhong_2018,hoch2021boson} or bulk optics \cite{wang2019boson} used for BosonSampling experiments \cite{brod2019photonic}. Other experimental aspects such as the impact of the degree of multi-photon distinguishability and exponentially-scaling photon losses on expressive power are subject for future research. 

While this article investigated the expressive power of the linear QPCs, the trainability and generalization power of the linear QPCs remains an open question. Apart from the gradient-free method used in this article for QML model training, gradient-based methods with analytical gradient \cite{schuld2019evaluating,banchi2021measuring,wierichs2022general} can potentially boost the training speed. However, current analytical gradient evaluation methods only apply to the photonic circuit with 1-photon input Fock state \cite{kerenidis2021classical} or continuous variable quantum photonic systems \cite{banchi2020training,miatto2020fast,yao2021fast,yao2022natural}. Hence, more research needs to be done to find the analytical gradient for the quantum photonic circuits with general input Fock state, which requires the differentiation of the permanents of the transfer matrix. It will also be interesting to see the effect of different input states, i.e. coherent states and squeezed states on the expressive power, trainability, and generalization power of the linear QPCs~\cite{afek2010high}. It will be interesting to further explore the translation of ideas between classical and quantum photonic circuits for machine learning.

\begin{acknowledgments}

This research was supported by the National Research Foundation, Prime Minister's Office, Singapore, the Ministry of Education, Singapore under the Research Centres of Excellence programme, and the Polisimulator project co-financed by Greece and the EU Regional Development Fund.

\end{acknowledgments}

\appendix
\begin{figure*}
\includegraphics[width=175mm,scale=0.5]{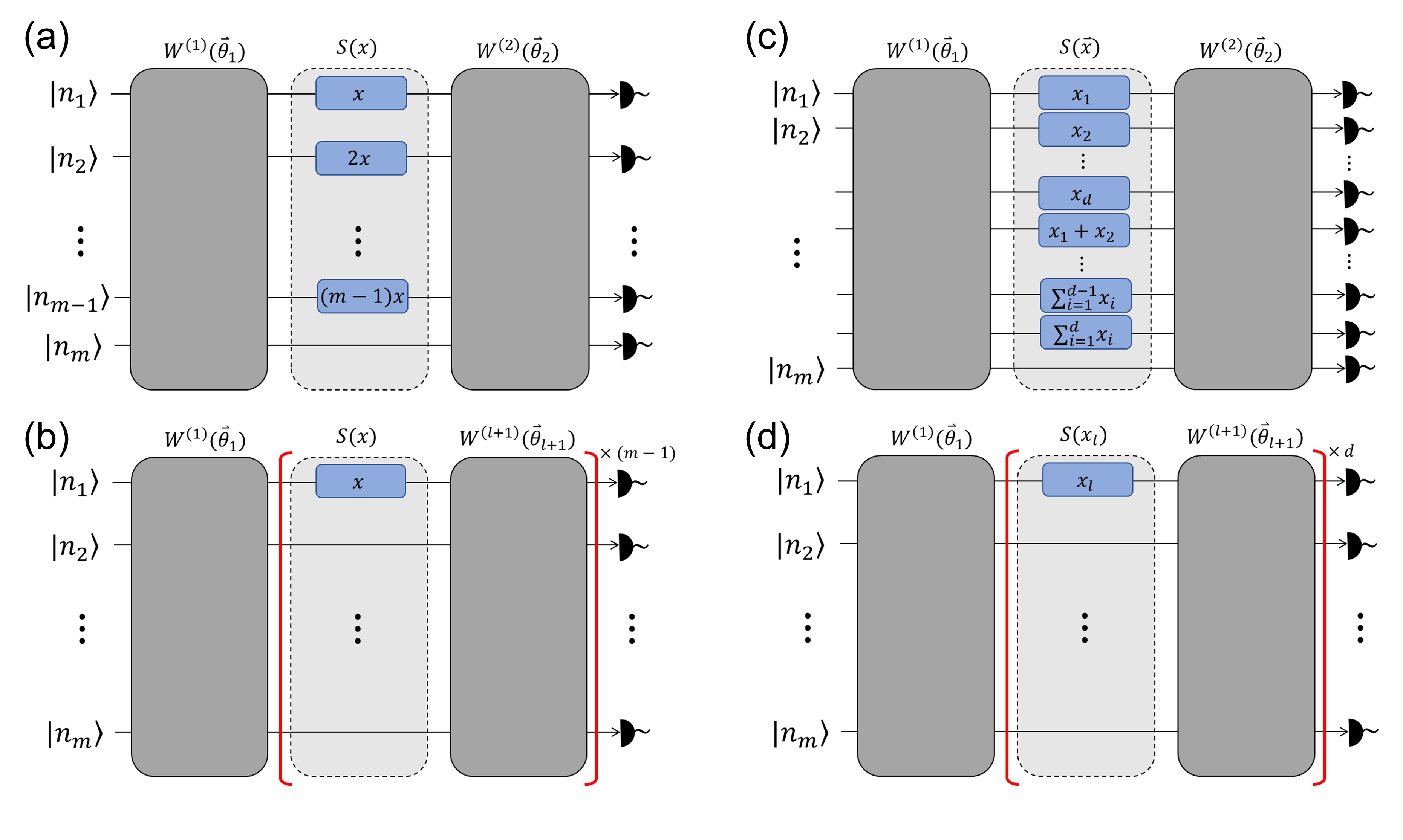}
\caption{\label{fig:appendix-Full_encoding_scheme_1DnMultiD} (a) Series encoding scheme that utilised all spatial mode within the data encoding block and maximize the photon-number dependent expressive power of the linear QPCs. (b) Parallel encoding scheme that generates the same set of frequency spectrum as in (a). (c) Series encoding scheme and (d) parallel encoding scheme that generate full frequency spectrum for $d$-dimensional Fourier series. The former demands $2^d - 1$ data encoding phase shifters while the later requires only $d$ phase shifters distributed equally among $d$ data encoding blocks.}
\end{figure*}

\section{\label{app:Encoding_scheme_1DFS}General encoding scheme for 1D frequency spectrum}
In Sec.\ref{subsubsec:n-dependent-expressivity}, we have derived the frequency spectrum for linear QPCs with single data encoding block that consists only one data encoding phase shifter. In this section, we broaden the frequency spectrum by adding more data encoding phase shifters into the same layer (series encoding) and different layers (parallel encoding). In the series encoding scheme [Fig.\ref{fig:appendix-Full_encoding_scheme_1DnMultiD}(a)], we consider linear QPCs with single data encoding block that consists of $m-1$ phase shifters, i.e. the highest possible number of phase shifters that can be placed within a data encoding block. For parallel encoding scheme [Fig.\ref{fig:appendix-Full_encoding_scheme_1DnMultiD}(b)], the $m-1$ phase shifters are equally distributed among $m-1$ data encoding blocks. One could consider different combinations of phase shifters in each layers and the expressive power will change accordingly. 

As shown in Sec.\ref{subsubsec:n-dependent-expressivity}, the size of the frequency spectrum of a $m$ mode linear QPC with one data encoding phase shifter is given by $D_{(n,1,1)} = n$, where $D_{(n,L,q)}$ (two additional subscripts are added for clarity) denotes the size of frequency spectrum realizable by linear QPCs with $n$ input photons and $L$ data encoding blocks, each block consists of $q$ data encoding phase shifters. For series encoding ($L=1$), we can place one data encoding phase shifter per mode on the first $m-1$ mode, each encodes phase proportional to its mode number [Fig.\ref{fig:appendix-Full_encoding_scheme_1DnMultiD}(a)], i.e. $i·x$ phase shift with $i$ denotes the mode number. The range of phases that could pick-up by $n$ photon is $[0,(m-1)n]$, where the lower (upper) bound is obtained when all photon passes through the last (second last) mode. Hence, the size of the frequency spectrum $D_{(n,1,m-1)}$ is $(m-1)n$. Identical range of phases is also apply to the parallel encoding scheme, where the lower (upper) bound is achieved when none (all) of the photon passes through the first mode on each layer, thus, $D_{(n,m-1,1)} = (m-1)n$.

\section{Encoding scheme to generate full frequency spectrum for multi-dimensional Fourier series}
\label{app:Encoding_scheme_multiDFS}

In this section, we will introduce the series and parallel encoding schemes that can generate a full frequency spectrum for multi-dimensional Fourier series. For series encoding scheme [Fig.\ref{fig:appendix-Full_encoding_scheme_1DnMultiD}(c)], one would need $2^d-1$ phase shifters to encode the positive phases of $d$-dimensional degree 1 Fourier series, i.e. $(\{\sum_{r = i}^d r_i x_i | r_1, r_2, ..., r_d \in \{0,1\}\}  \backslash \{0\} )$ .  For example, one can use 7 phase shifters to encode $\{x_1,x_2,x_3,x_1+x_2,x_1+x_3,x_2+x_3,x_1+x_2+x_3\}$. Then, the frequency spectrum of $d$ dimensional degree $n$ Fourier series, i.e. $\Omega^{(d)}_n = (-\bm{\omega}^{(n)},0, \bm{\omega}^{(n)} )$ with $\bm{\omega}^{(n)} = (\omega^{(n)}_1,\omega^{(n)}_2,...,\omega^{(n)}_d)$ and $\omega^{(n)}_i \in \{0,1,...,n\}$ can be generated using $n$ input photons. On the other hand, the same set of frequency spectrum can be generated using $d$ data encoding blocks, each consists of one data encoding phase shifter that encodes one data feature. [Fig.\ref{fig:appendix-Full_encoding_scheme_1DnMultiD}(d)].

\bibliography{bibliography.bib}
\end{document}